\documentclass[prd,aps,showpacs,nofootinbib,twocolumn,amssymb,amsmath,amsfonts]{revtex4}
\usepackage{latexsym}
\usepackage[latin1]{inputenc}

\usepackage{slashed}
\usepackage{dcolumn}% Align table columns on decimal point
\usepackage[dvips]{graphicx}
\usepackage{color}

\newcommand{\be}{\begin{equation}}
\newcommand{\ee}{\end{equation}}
\newcommand{\ba}{\begin{eqnarray}}
\newcommand{\ea}{\end{eqnarray}}

\newcommand{\Rvec}{{\bf R}}
\newcommand{\rvec}{{\bf r}}
\newcommand{\Pvec}{{\bf P}}
\newcommand{\pvec}{{\bf p}}
\newcommand{\Avec}{{\bf A}}
\newcommand{\Bvec}{{\bf B}}
\newcommand{\ma}{\mathcal{A}}
\newcommand{\pa}{\partial}

\newcommand{\nn}{\nonumber}

\begin{document}

\title{Chiral transport equation from the quantum Dirac Hamiltonian \\ and the on-shell effective field theory}
\author{Cristina Manuel}
\affiliation{Institut de Ci\`encies de l'Espai (IEEC/CSIC), Campus Universitat Aut\`onoma
de Barcelona, Facultat de Ci\`encies, Torre C5, E-08193 Bellaterra, Spain}
\author{Juan M. Torres-Rincon}
\affiliation{Institut de Ci\`encies de l'Espai (IEEC/CSIC), Campus Universitat Aut\`onoma
de Barcelona, Facultat de Ci\`encies, Torre C5, E-08193 Bellaterra, Spain}
\affiliation{Subatech, UMR 6457, IN2P3/CNRS, Universit\'e de Nantes, \'Ecole de Mines de Nantes, 4 rue Alfred Kastler 44307,
Nantes, France}


\begin{abstract}
We derive the relativistic chiral transport equation for massless fermions and antifermions by performing a semiclassical Foldy-Wouthuysen diagonalization of the quantum Dirac Hamiltonian. The Berry connection naturally
emerges in the diagonalization process to modify the classical equations of motion of a fermion in an electromagnetic field. We also see that the fermion and antifermion dispersion
relations are corrected at first order in the Planck constant by the Berry curvature, as previously derived by Son and Yamamoto for the particular case of vanishing temperature. 
Our approach does not require knowledge of the state of the system, and thus it can also be applied at high temperature. 
 We provide support for our result by an alternative computation using an effective field theory for fermions and antifermions: the on-shell effective field theory.
In this  formalism, the off-shell fermionic modes are integrated out to generate an effective Lagrangian for the quasi-on-shell fermions/antifermions. The dispersion relation at leading order exactly matches the result from the semiclassical diagonalization. 
From the transport equation, we explicitly show how the axial and gauge anomalies are not modified at finite temperature and density despite the incorporation of the new dispersion relation into the distribution function.
\end{abstract}
\maketitle

\section{Introduction}

Symmetries in classical physical systems are not always preserved when the laws of quantum
mechanics are taken into account.  One remarkable example of this fact is given by the quantum chiral  anomaly, which
explains  the  decay of the neutral pion into two photons~\cite{Adler,BellJackiw}. While the implications in particle physics of the chiral anomaly have
been thoroughly studied, its macroscopic consequences have not been enough explored.
Only recently has it been pointed out that the relativistic hydrodynamical equations,
which are the expressions of the conservation laws of a system, 
would be affected by the chiral anomaly~\cite{Son:2009tf}. In particular, all these studies have triggered interest in studying macroscopical effects of the
quantum chiral anomaly in heavy ion colliders (see the reviews in Refs.~\cite{Kharzeev:2013ffa,Landsteiner:2012kd} and references therein).

In a series of references~\cite{Son:2012wh,Stephanov:2012ki,Son:2012zy,Chen:2012ca} 
 it has been shown how the chiral anomaly equation could be derived
from classical transport theory in systems with well-defined Fermi surfaces. These references have
shown that if a Berry connection is included in  the particle action of a relativistic fermion,  and  noncanonical
Poisson brackets are taken into account, the associated classical transport approach reproduces the triangle anomaly
present in  quantum field theory. The present work is in part motivated by the fact that
while Refs.~\cite{Son:2012wh,Stephanov:2012ki,Son:2012zy,Chen:2012ca} seem to describe
the same transport equation, they do not. The different proposals differ in the form of
the  fermionic dispersion law, and also in explicit terms that appear in the kinetic equation.
 Thus, they lead to a different dynamical evolution of the
fermion distribution function. Only the authors of Ref.~\cite{Son:2012zy} made a derivation
of some (but not all) the terms that appear in their proposed transport equation, starting
with an effective field theory valid only for fermion modes close to a well-defined Fermi surface.
The proof, thus, was only strictly valid at zero temperature $T=0$, where antiparticles degrees of freedom
are integrated out. In Ref.~\cite{Manuel:2013zaa}, a consistency check of the proposal of Ref.~\cite{Son:2012zy} 
was made. It was then shown that if the same equation was used for particles and antiparticles at 
finite $T$  some anomalous Feynman diagrams taken in the
so-called hard thermal loop approximation could be reproduced~\cite{Laine:2005bt}.

In this manuscript we present a derivation of the chiral transport equation
of Son and Yamamoto. It is based on deriving quantum corrections to the
classical equations of motion of a charged massless fermion in the background of
an electromagnetic field. In order to do so, we start with the quantum Dirac Hamiltonian,
and perform a Foldy-Wouthysen diagonalization~\cite{Foldy:1949wa} in powers of the Planck constant 
$\hbar$, following the methods and results of Ref.~\cite{Gosselin:2006ht}, see also~\cite{Bliokh:2005nk}.
The rotated position and momentum operators can be obtained at the same order in $\hbar$ as well,
and their associated commutators turn out to be noncanonical. 
We then take the diagonalized Hamiltonian at order $\hbar$ and treat it classically, together with the Poisson brackets (PB) that
can be deduced from the quantum commutators of the theory.
In this way we can reproduce the transport equation first proposed by Son and Yamamoto, both for particles and
for antiparticles, and without any reference to the state of the system, and in particular to having
a well-defined Fermi surface. 
We can  understand the quantum origin of the noncanonical PB structure of the theory, and
that the fermion and antifermion dispersion laws are affected by the Berry connection at the order
of accuracy at which one is working.

A complementary proof can also be presented at a quantum field theory level.
 Using effective field theory techniques, it is also possible to disentangle the particle and antiparticle
degrees of freedom associated with the Dirac field. We propose what we call 
on-shell effective field theory, which aims to describe the almost on-shell propagation
of massless fermions and antifermions. Using these methods we check the fermion and
antifermion dispersion laws obtained through the semiclassical diagonalization of the
Dirac Hamiltonian. We then point out that by following the same methods and reasoning
as Son and Yamamoto, one could in principle derive from our effective field theory their chiral transport equation
without references to having a Fermi surface.

This paper is structured as follows. In Sec.~\ref{sec-diagonalization} we review the
process of diagonalization of the Dirac Hamiltonian at order $\mathcal{O} (\hbar)$, and how
the Berry connections modify the classical Hamiltonian, and change the commutators of
the space and momentum operators for a massive fermion. In Sec.~\ref{transp-sec} we use
the Hamiltonian obtained previously to derive semiclassical corrections in classical transport
theory for massless fermions, and thus show how to derive the transport approach of
Son and Yamamoto. In Sec.~\ref{EFT-sec} we derive an effective field theory valid to
describe processes dominated by  massless quasi-on-shell fermions, which we use to check the
semiclassical corrections to the fermion and antifermion dispersion laws.
We use natural units $c=k_B= \hbar= 1$ in the manuscript, but keep $\hbar$ explicitly in
Secs.~\ref{sec-diagonalization} and \ref{transp-sec} to show all the $\hbar$ corrections to the classical
dynamics that have to be consistently kept. An appendix is devoted to showing the details of
the derivation of the chiral anomaly equation for a relativistic plasma.

\section{Semiclassical diagonalization}
\label{sec-diagonalization}

We review in this section the semiclassical  diagonalization of the quantum Dirac Hamiltonian to order $\mathcal{O} (\hbar)$ performed in Ref.~\cite{Gosselin:2006ht}. After
checking the results of Ref.~\cite{Gosselin:2006ht} we have detected some relevant typos in its final equations that are important for our present discussion, and we will
point them out below.

Let us consider a relativistic fermion of mass $m$ in an electromagnetic field. The 
Dirac Hamiltonian reads
\be H_0 (\Pvec,\Rvec) = \alpha \cdot (\Pvec- e \Avec(\Rvec)) + \beta m + e A_0(\Rvec) \ , \ee 
where in the Dirac basis,
\be \alpha_k=\left( \begin{array}{cc}
0 & \sigma_k \\
\sigma_k & 0 \label{eq:alpha}

\end{array} \right) ,  \quad \beta=\left( \begin{array}{cc}
I & 0  \\
 0 &-I 
\end{array} \right) \ , \ee
and $\sigma_k$ and $I$ are the Pauli and unit  $2 \times 2$ matrices, respectively.
 Here $A_\mu = (A_0 , \Avec)$ is the vector gauge field potential.
For simplicity, we will assume that it does not depend on time.
This Hamiltonian is explicitly nondiagonal in this  basis due to the presence of the Dirac matrices ${\bf \alpha}$.

The classical Hamiltonian, i.e. assuming that $\Rvec$ commutes with $\Pvec$, can be diagonalized by performing a Foldy-Woythuysen transformation~\cite{Foldy:1949wa} with the help of the unitary matrix
\be U (\Pvec,\Rvec) = \frac{E+m+\beta \alpha \cdot (\Pvec- e\Avec (\Rvec))}{\sqrt{2E(E+m)}} \ , \ee
with $E \equiv \sqrt{(\Pvec- e\Avec (\Rvec))^2+m^2}$.
The resulting Hamiltonian $H_D$
\be H_D = U H_0 U^\dag = \beta E  + eA_0(\Rvec)
\ee
is diagonal by blocks in Dirac space, with the upper-left block corresponding to particles and the bottom-right to antiparticles.

When one considers the noncommuting character of the  canonical variables --that is $[R_i,P_j]=i\hbar \delta_{ij}$-- it is still possible to perform a perturbative diagonalization in powers of $1/m$.
This diagonalization allows to reproduce the nonrelativistic Pauli Hamiltonian, as well as higher relativistic corrections~\cite{Foldy:1949wa}.  Although not so widely known,
 it is also possible to perform a perturbative diagonalization in powers of the Planck constant $\hbar$~\cite{Bliokh:2005nk,Gosselin:2006ht}. This last diagonalization is the only  one possible  for the 
massless fermions we will be considering in the next section of this manuscript.  Here we will briefly describe how to achieve this separation between particle and antiparticle degrees of freedom at order $\mathcal{O} (\hbar)$~\cite{Gosselin:2006ht}.

It is evident that at order  $\mathcal{O} (\hbar^0)$, when the canonical variables commute, the matrix $U$ diagonalizes the Dirac Hamiltonian. If we consider that the canonical variables do not commute, then it is necessary to choose a
prescription for the operators where products of the canonical variables appear. One possible choice is to symmetrize all operators that depend on ${\bf P}$ and ${\bf R}$. By symmetrization one means that each operator has to be written in a form where
half of all the powers of ${\bf P}$ have been put  on the left, and the other half on the right.  
 
After the symmetrization process, the matrix $U({\bf P},{\bf R})$ does not remain unitary. Thus, one rather uses the matrix $U({\bf P},{\bf R})+ XU({\bf P},{\bf R})$, where $X$ is the contribution that allows to preserve unitarity at order  $\mathcal{O} (\hbar)$.
This matrix can be explicitly computed and reads~\cite{Gosselin:2006ht} 
\be
X = \frac{i}{4 \hbar} [\ma_{P^i},\ma_{R_i}] \ ,
\ee
written in terms of the Berry connections
 \ba 
 \ma_{R_i} &=& i \hbar U \nabla_{P_i} U^\dag   \ , \\
 \ma_{P_i} &=& -i \hbar U \nabla_{R_i} U^\dag \ . \ea
 
The reason why the above functions are called Berry connections comes from the fact that they give rise to the same  Berry curvatures~\cite{Berry:1984jv}  that can be computed starting from the momentum eigenstates of the Dirac Hamiltonian.
An explicit evaluation of these functions from the matrix $U$ gives
 
 \begin{widetext}
 \be \label{eq:berryr} \ma_{R_i} = \hbar \frac{-i \alpha \cdot (\Pvec - e\Avec)(P_i - eA_i) \beta - i \beta E (E+m) \alpha_i - E [ {\bf \Sigma} \times (\Pvec - e\Avec)]_i}{2E^2 (E+m)} \ , \ee
 \end{widetext}
 where
 \be \Sigma_k = \left( \begin{array}{cc}
 \sigma_k & 0 \\
0 & \sigma_k 
\end{array} \right) \  \ee
is the spin matrix and
 \be \label{eq:berryp}
 \ma_{P^i}=  e\ \nabla_{R^i} A_k (\Rvec) \ma_{R^k} \ .
 \ee

Note that this diagonalization is respectful of the gauge symmetry. While the  Berry connection $\ma_{R_i}$ only depends on $\Pvec - e\Avec(\Rvec)$, operating as a covariant derivative when acting on the fermion wave function,
the connection $ \ma_{P_i} $ explicitly breaks gauge invariance, as it depends on $\nabla_{R^i} A_k$. However, it is easy to prove that $X$, and thus $U + X U$,  has the gauge covariant properties one requires to preserve the gauge symmetry.

After the unitary transformation, the new Hamiltonian still includes nondiagonal terms of order ${\cal O} (\hbar)$. However, as proven in Ref.~\cite{Gosselin:2006ht}, the true diagonalized Hamiltonian coincides with the projection on the diagonal of the transformed Hamiltonian at this stage.
After a straightforward but tedious computation, one arrives to an almost diagonal Hamiltonian, which when projected on the diagonal gives 
\begin{widetext}
\be H_D = \beta E (\rvec,\pvec) + \frac{i}{2\hbar} \ P \{ \ [ \beta E(\rvec,\pvec),\ma_{R_i} ] \ma_{P^i} - [ \beta E (\rvec,\pvec), \ma_{P^i}] \ma_{R_i} \ \} + eA_0(\rvec) \ , \ee
\end{widetext}
where $P$ is the projection over the diagonal operator and $E (\rvec,\pvec)$ is
\be E(\rvec,\pvec)= \sqrt{(\pvec-e\Avec(\rvec))^2+m^2} 
\ . \ee
Working out the commutators and projecting into the diagonal we find
\be
 \label{eq:ham} 
H_D =  \beta E -  \beta \frac{e\hbar {\bf  \Sigma} \cdot {\bf B}}{2E} - \beta \frac{e {\bf L} \cdot {\bf B}}{E} +e A_0(\rvec) \ , \ee
where an intrinsic angular momentum, which takes into account spin-orbit effects, 
 has been introduced as
 \be {\bf L} = \tilde{\bf p} \times P ({\bf \ma_{R}})  =  \hbar \ \frac{ \tilde{\bf p} \times (\tilde{\bf p} \times {\bf \Sigma})}{2E(E+m)} \ ,
\ee
where $\tilde{\bf p}= \pvec - e\Avec(\rvec)$ is the kinetic momentum.

Equation~(\ref{eq:ham}) is 
the same Hamiltonian that appears in Ref.~\cite{Gosselin:2006ht}, although we note a typo in their Eq.~(49).
It also coincides with the result of Ref.~\cite{Bliokh:2005nk}.

We have already written the Hamiltonian in the new basis in terms of the rotated position and momentum operators, which at order $\mathcal{O} (\hbar)$
read 
\ba \label{eq:newr}
 \rvec & = & P[U({\bf P},{\bf R})\, {\bf R} \ U^\dagger({\bf P},{\bf R})] 
 = \Rvec + P (\ma_{R}) \ , \\ 
 %= \Rvec + A_{R} \ , \\
\label{eq:newp}
 \pvec & = & P[U({\bf P},{\bf R}) \,{\bf P} \ U^\dagger({\bf P},{\bf R})] 
 =  \Pvec + P (\ma_{P})
 %= \Pvec + A_{P} 
 \ . \ea

It is possible to check that the new Hamiltonian preserves gauge invariance at order $\mathcal{O} (\hbar)$, 
as the new kinetic momentum $\tilde{\bf p}= \pvec - e\Avec(\rvec)$ can be checked to preserve the gauge symmetry  if one expands $\Avec(\rvec)$ at that order.

Not surprisingly,  the commutation relations of the new variables turn out to be noncanonical. Their basic commutators can be derived from  Eqs.~(\ref{eq:newr},\ref{eq:newp})  as
\ba 
[r_i,r_j] &=& [R_i, P(\ma_{R_j})] - [R_j,P(\ma_{R_i})] + [P(\ma_{R_i}),P(\ma_{R_j})] \ ,  \nn \\
\left[p_i,p_j\right] &=& [P_i, P(\ma_{P_j})] - [P_j, P(\ma_{P_i})] + [P(\ma_{P_i}),P( \ma_{P_j})] \ , \nn \\
\left[r_i,p_j\right] &=& i \hbar \delta_{ij} + [R_i,P(\ma_{P_j})] - [P_j, P(\ma_{R_i})]  \nn \\
&+ & [P(\ma_{R_i}),P(\ma_{P_j})] \label{eq:comms} \ . 
\ea

The explicit evaluation of the above commutators gives
\be
 [r_i, r_j]  =  i \hbar^2  G_{ij} = - i \hbar^2 \epsilon_{ijk} G_k \ ,  
\ee
where
\be
\label{field-p}
{\bf G}(\tilde{\bf p}) = \frac{1}{2 E^3} \left( m {\bf \Sigma} + \frac{ (\bf{\Sigma} \cdot  \tilde{\bf p})\tilde{p} }{E+m} \right) \  .
\ee
Here
$ G_{ij}$   can be interpreted as a field strength tensor in momentum space~\footnote{It is interesting to note here
that while the Berry connection for massless fermions transforms under $U(1) \times U(1)$ rotations, in the massive case
the Berry connections are non-Abelian, and transforming under the $SU(2)$ group~\cite{Jackiw:1987vi} (see also the coincident 
results in Ref.~\cite{Chang:2008zza}, where the Berry curvature vector defined there corresponds to our $-\hbar^2 {\bf G}$).}. 

The remaining commutators are better expressed in terms of the kinetic momentum as
\ba 
{[\tilde{p}_i,\tilde{p}_j]} &=&  i e \hbar F_{ij} + i e^2 \hbar^2 F_{ik} F_{jm} G_{km}  \ , \\
{[r_i,\tilde{p}_j]} &=&   i \hbar \delta_{ij}  + i e \hbar^2 F_{jk} G_{ik} \ .
\ea
These relations coincide with those presented in  Ref.~\cite{Bliokh:2005nk}.

With the basic commutators one could derive the associated equations of motion. We will only write them in the next section for massless fermions.

\section{Semiclassical kinetic theory for massless fermions}
\label{transp-sec}

The Hamiltonian derived in Eq.~(\ref{eq:ham}) can be used to derive semiclassical corrections --that is, corrections of
order  $\mathcal{O} (\hbar)$ to the classical dynamics. In particular, it can be used to derive the first $\hbar$ corrections
to the well-known Vlasov transport equations. We will prove in this section that when we use the semiclassical Hamiltonian together
with the associated  noncanonical variables in the massless case we reproduce the chiral kinetic theory of Ref.~\cite{Son:2012zy}. We should stress
here that for massive fermions the same method and reasoning could be used to obtain their corresponding transport equations, although such a
case will be treated elsewhere (see for instance the recent results of Ref.~\cite{Chen:2013iga}, but where the fermion dispersion relation was taken at ${\cal O} (h^0)$).

Several drastic simplifications occur in the  dynamics in the massless case. We first note that when $m=0$ the Hamiltonian Eq.~(\ref{eq:ham}) can be expressed as
\be
 \label{eq:hamiltonian} 
 H_D = \beta  p \left( 1 - e \hbar \, \lambda {\bf \Omega} \cdot \Bvec \right) + eA_0(\rvec) \ , \ee
where we have defined the Berry curvature
\be 
{\bf \Omega}=  \frac{\bf p}{2 p^3} 
 \ee
and the helicity operator,
\be
 \lambda = \frac{\Sigma \cdot \bf p }{p} \ , \ee
which can take values $\pm 1$. Note that we have renamed in this section the kinetic momentum
$\tilde{\bf p}$ as ${\bf p}$ to simplify the notation.
Note also that for $m=0$, the field in Eq.~(\ref{field-p}) is now written as ${\bf G} = \lambda {\bf \Omega}$.

From the above Hamiltonian, we can immediately read the dispersion relations.
Assuming $A_0 =0$, the dispersion relation for the particle reads
\be \label{eq:disp1} \epsilon^+_{\bf p}= p \left( 1 - e\hbar  \, \lambda \frac{\Bvec \cdot \pvec}{2p^3} \right) \ , \ee
while  for the antiparticle 
\be \label{eq:disp2} \epsilon^-_{\bf p}= -p \left( 1 - e\hbar \, \lambda \frac{\Bvec \cdot \pvec}{2p^3} \right) \ . \ee

In the semiclassical diagonalization we carried out, in an $\hbar$ expansion, one assumes that quantum corrections
to the classical physics should be small. Thus, the validity of our results requires that  $e \hbar \Bvec \cdot
\pvec/p^3 \ll 1$, as otherwise one should keep more corrections in the diagonalization procedure.

The dispersion relations (\ref{eq:disp1},\ref{eq:disp2}) have been used in previous works --see Refs.~\cite{Son:2012zy,Manuel:2013zaa,Akamatsu:2014yza}--
although only the particle dispersion relation has been derived for fermions close to the Fermi surface at temperature $T=0$ using effective field
theory methods~\cite{Son:2012zy}. Here we see that the dispersion relations used in these references can naturally be
derived after performing the semiclassical diagonalization of the massless quantum Dirac Hamiltonian.

If we want to compute $\hbar$ corrections to the classical  equations of motion of a  fermion in 
an electromagnetic field, we need the Poisson brackets (PB) of the  associated position and momentum variables. These can
be deduced from the quantum commutators, through the standard rule $\frac{1}{i \hbar} [ \cdot  , \cdot ] \rightarrow \{ \cdot  , \cdot \}_{\rm PB}$.  
In the massless case, and from the quantum commutators obtained in the previous section one finds
 \ba
 \{r_i, r_j \}_{\rm PB}  & \approx & - \hbar \epsilon_{ijk} \, \lambda \Omega_k \ ,  \\
 \{p_i, p_j \}_{\rm PB} & \approx & e \,  \epsilon_{ijk} B_k  \ , \\
 \{ r_i,p_j \}_{\rm PB} & \approx & \delta_{ij} \ ,
 \ea
where we have only kept the leading $\hbar$ corrections in the coordinate PB, and  have neglected other quantum corrections
in front of the classical values otherwise.

With the Hamiltonian in Eq.~(\ref{eq:hamiltonian}) and the PBs above one obtains the classical equations of motion for a right-handed fermion ($\lambda = 1$) 
\ba
\dot{\bf p}& = & - \frac{\partial \epsilon^+_{\bf p}}{\partial \bf r} + e ({\bf E} + \dot {\bf r}\times {\bf B}) \ , \\
\dot{\bf r} & = & \frac{\partial \epsilon^+_{\bf p}}{\partial \bf p} - \hbar  (\dot {\bf p}\times {\bf \Omega }) \ .
\ea
Substituting one equation into the other, one then gets
\ba
\label{dr/dt}
 (1 + e \hbar {\bf B}\cdot {\bf \Omega}) \dot{\bf r} & = & \tilde {\bf v} + e  \hbar ( \tilde{\bf E}  \times {\bf \Omega})
 + e \hbar {\bf B} (\tilde{\bf v} \cdot {\bf \Omega})  , \\
 \label{dp/dt}
 (1 + e \hbar {\bf B}\cdot {\bf \Omega}) \dot{\bf p} & = &  e ( \tilde{\bf E}  + \tilde {\bf v}\times {\bf B}) + e^2 \hbar {\bf \Omega}  ( \tilde{\bf E}  \cdot {\bf B}) ,
  \ea
 where we have defined 
 \ba
 \tilde{\bf E} & = & {\bf E}- \frac 1e \frac{\partial \epsilon^+_{\bf p}}{\partial \bf r}  \ ,
 \\
\tilde {\bf v} & =&  \frac{\partial \epsilon^+_{\bf p}}{\partial {\bf p}} \ .
\ea
Similar equations could be found for left-handed particles, or for antiparticles of any helicity.

Note that these equations are essentially the same as those first written in Ref.~\cite{Stephanov:2012ki} (see also Ref.~\cite{Chen:2012ca}),
except for the fact that in these references the fermion dispersion relation was taken as  $\epsilon_p = p$. Here we see that this approximation 
is not correct if one wants to consistently take into account all the $\hbar$ corrections to the classical dynamics.
Furthermore, we have checked in Ref.~\cite{Manuel:2013zaa} that the anomalous hard thermal loop diagrams that appear in a
quantum field theory computation can
 {\it only}  be reproduced if the $\hbar$ corrections to the fermion and antifermion dispersion laws are taken into account.

It is now possible to compute semiclassical corrections to the transport equation. Using Liouville's theorem, one can
evaluate the dynamical evolution of the one-particle distribution function $f_p$, and find the kinetic equation 
 \be \frac{df_p}{dt} = \frac{\pa f_p}{\pa t} +  \dot{ {\bf r}} \cdot \frac{\pa f_p}{\pa {\bf r}} +  \dot{ \bf{p}} \cdot \frac{\pa f_p}{\pa {\bf p}} =0 \ , \ee
which can be expressed for $\lambda = 1$ as
\begin{widetext}
 \be
 \label{chiraltreq}
  \frac{\pa f_p}{\pa t} + \left( 1+ e \hbar \Bvec \cdot {\bf \Omega}\right)^{-1} \left\{ \left[  \tilde {\bf v} + e  \hbar \ \tilde{\bf E}  \times {\bf \Omega} 
 + e \hbar \ {\bf B} (\tilde{\bf v} \cdot {\bf \Omega}) \right]  \cdot \frac{\pa f_p}{\pa {\bf r}} +  e \left[  \tilde{\bf E}  + \tilde {\bf v}\times {\bf B} + e  \hbar {\bf \Omega} \
  ( \tilde{\bf E}  \cdot {\bf B}) \right] \cdot \frac{\pa f_p}{\pa {\bf p}} \right\} =0 \ . \ee
  \end{widetext}
  This equation agrees {\it exactly} with the chiral kinetic equation first proposed by Son and Yamamoto in Ref~\cite{Son:2012zy}. It also agrees with the
  transport equation of a Bloch electron in a solid previously derived in Ref.~\cite{Duval:2005vn}. Note that in the
  $\hbar \rightarrow 0$ limit, it reduces to the standard kinetic equation of a charged particle in the presence of  an electromagnetic field.

Given the noncanonical character of the ${\bf r}, {\bf p}$ variables, the invariant measure of the phase-space integration has to be modified~\cite{Morrison:1998zz}. It is  given by
  $\left( 1+ e \hbar \Bvec \cdot  {\bf \Omega}\right) d^3r d^3 p/(2 \pi \hbar)^3$~\cite{Duval:2005vn}.
  Then the particle density associated to right-handed fermions  reads
  \be n = \int \frac{d^3p}{(2\pi \hbar)^3} (1 + e \hbar \Bvec \cdot {\bf \Omega}) f_p \ ,
   \ee 
 while the associated current can be expressed as~\cite{Son:2012zy} 
\begin{widetext}
\be {\bf j} = - \int \frac{d^3 p}{(2\pi \hbar)^3} \left[ \epsilon_p \frac{\pa f_p}{\pa \pvec}  + e \hbar {\bf \Omega} \cdot \frac{\pa f_p}{\pa \pvec} \epsilon_p \Bvec +  \hbar \epsilon_p  {\bf \Omega} \times \frac{\pa f_p}{\pa \rvec}  - e\hbar f_p {\bf E} \times {\bf \Omega} \right] \ . \ee
\end{widetext}

After integrating the kinetic equation over the (invariant) momentum measure one then obtains
\be
\label{pre-anomaly}
 \frac{\pa n}{\pa t} + \nabla \cdot {\bf j} = - e^2 \hbar \int \frac{d^3p}{(2\pi \hbar)^3} \left( {\bf \Omega} \cdot \frac{\pa f_p}{\pa \pvec} \right) {\bf E} \cdot {\bf B} \ . \ee
Written in this form, one sees the clear quantum origin of the nonconservation of the current, because if the quantum corrections to the classical equations
of motion were neglected, the current  would be  conserved.

From Eq.~(\ref{pre-anomaly}) one can deduce the same  quantum chiral anomaly equation of quantum field theory for an equilibrium relativistic plasma
if we take into account the different contributions of particles and antiparticles of different helicities that are present in the plasma.
In Ref.~\cite{Manuel:2013zaa} (see also the footnote of Ref.~\cite{Stephanov:2012ki})
 we have computed this integral for the equilibrium one-particle distribution function to reproduce the equation of the chiral anomaly, and check that it does not receive thermal corrections.
  However, there the  fermion dispersion relation was taken at order ${\cal O} (\hbar^0)$, which is not completely consistent, as
 $\hbar$ corrections also affect   the fermion dispersion law.
 It is still possible to show that the chiral anomaly equation is not affected if the $\hbar$ correction to the fermion dispersion law
 is taken into account.
 In order to do so, we proceed as in Ref.~\cite{Manuel:2013zaa}, and define a sphere of radius $\Delta$ centered at ${\bf p} =0$ to evaluate Eq.~(\ref{pre-anomaly}).
The nonzero divergence of  the current can then be interpreted as arising from the flux of fermions that cross the surface $S^2(\Delta)$ defined by that sphere.  Then taking the $\Delta \rightarrow 0$ limit
 \be
 \label{eq:divj} \frac{\pa n}{\pa t} + \nabla \cdot {\bf j} = e^2 {\hbar} \lim_{\Delta \rightarrow 0} \int_{S^2(\Delta)} \frac{d {\bf S}}{(2\pi \hbar)^3} \cdot {\bf \Omega} \ f_p {\bf E}  \cdot {\bf B}  \ .   
 \ee

In thermal equilibrium, the distribution functions for right- and left-handed fermions and antifermions read
\ba \label{eq:distri1}f^{R,L}_p &=& \frac{1}{\exp\left[ \frac{1}{T} \left( p  \mp  e\hbar \frac{{\bf B} \cdot \pvec}{2p^2} - \mu_{R,L} \right) \right] +1} \ , \\
\label{eq:distri2} \bar{f}^{L,R}_p &=& \frac{1}{\exp\left[ \frac{1}{T} \left( p  \pm  e\hbar \frac{{\bf B} \cdot \pvec}{2p^2} + \mu_{R,L} \right) \right] +1} \ . \ea

The computation of the chiral anomaly equation can be exactly done with the new dispersion relation. Details are given in App.~\ref{app:integral}. The divergence of the axial current $j^\mu_A = j^\mu_R -j^\mu_L$ 
is easily derived to obtain
\be \pa_\mu j^\mu_A = \frac{e^2}{2\pi^2 \hbar^2} {\bf E} \cdot {\bf B}  = \frac{2 \alpha}{\pi \hbar}  {\bf E} \cdot {\bf B} 
\ , \ee
which agrees exactly with the same result if we would have ignored the quantum corrections to the fermion dispersion law, see Ref.~\cite{Manuel:2013zaa}.
The same calculation reveals that these corrections do not affect the gauge anomaly, that is, that the electromagnetic (or any vectorial) current is conserved.

\section{On-Shell Effective Field Theory}
\label{EFT-sec}

The semiclassical diagonalization performed in Sec.~\ref{sec-diagonalization} should have a counterpart in a quantum field theoretical framework.
In fact, it is possible to disentangle the particle and antiparticle degrees of freedom using effective field theory (EFT) techniques. 
In this section we introduce an EFT for almost on-shell massless fermions and antifermions, which is basically obtained by integrating out
all the off-shell modes. Therefore, we call it on-shell effective field theory (OSEFT). It is aimed to be used for the description of processes dominated
by on-shell fermions in an ultrarelativistic plasma. In this work we will restrict ourselves to show that with this EFT one can have an independent check of
the particle and antiparticle dispersion relations written in Eqs.~(\ref{eq:disp1}, \ref{eq:disp2}), leaving further developments for a future work \cite{future}. We will also see that the methods of Ref.~\cite{Son:2012zy}
could also be used to derive the chiral transport equation without any reference to a well-defined Fermi surface.
We are inspired by many other EFTs, in particular, by the large energy effective theory (LEET), as formulated in Ref.~\cite{Charles:1998dr}, and by the high density effective theory (HDET) of
Refs.~\cite{Hong:1998tn,Hong:1999ru}, which is, however, only valid at finite density and $T=0$.
In order to simplify notation we will again work in natural units $\hbar =1$.

Let us consider a relativistic electromagnetic plasma close to thermal equilibrium, at temperature $T$, and composed by massless fermions.
The propagation of an on-shell fermion is in principle described by its energy $E= p$ and the four lightlike four-velocity $v^\mu= (1, {\bf v})$, where
${\bf v}$ is three-dimensional unit vector, in such a way that $p^\mu = p v^\mu$. The four-momentum of a fermion which is nearly on-shell can be written as
\be
q^\mu = p v^\mu + k^\mu \ ,
\ee
where $k^\mu$ is the residual momentum, i.e. the part of the momentum which makes $q^\mu$ off shell. 
Requiring that $q^2$ be small is equivalent to demanding that $v \cdot k \ll p$ and $k^2 \ll q^2$. These constraints
basically impose that the component of the three-momentum orthogonal to ${\bf v}$ is ${\bf k}_\perp \ll p$.
 
A similar decomposition of the momentum for almost on-shell antifermions can be done  as follows
\be
q^\mu = - p {\tilde v}^\mu + k^\mu \ ,
\ee
where ${\tilde v}^\mu = (1 , -{\bf v})$.

We define the particle/antiparticle projectors
\be
P_{\pm v} = \frac{ 1 \pm {\bf \alpha}\cdot {\bf v}}{2}  \ , \qquad   \alpha_i = \gamma_0   \gamma_i \ ,
\ee
where in the Dirac representation $\alpha_i$ is given by Eq.~(\ref{eq:alpha}). With the usual projector relations
\be
P^2_{\pm v} = P_{\pm v}   \ ,  \qquad P_{\pm v} P_{\mp v} = 0 \ , \qquad   P_{+v} + P_{-v} = 1 \ ,
\ee
we can write  the Dirac field  as
\ba
\label{factor-wf}
\psi_v(x) & = &   e^{- i p v \cdot x} \left ( P_{+v} \chi_{+v} (x) + P_{-v} H^1_{-v}(x) \right ) \\
&+&   e^{ i p {\tilde v} \cdot x} \left (  P_{-v} \xi_{-v} (x) + P_{+v} H^2_{+v}(x) \right ) \ ,
\nonumber
\ea
where we have factored out the dependence on $p$, so the particle/antiparticle fields appearing
in the right-hand side of Eq.~(\ref{factor-wf}) only depend on the residual momentum.

If we want to describe only fermions and antifermions with arbitrary velocity directions, the corresponding Lagrangian can then be written as
a sum over different velocities,
\be
{\cal L}_f = \sum_{\bf v} {\cal L}_{fv} \ ,
\ee
with
\be
{\cal L}_{fv} = \bar{\psi}_v  i\slashed{D} \psi_v =  \bar{\psi}_v (i \slashed{\partial} - e \slashed{A}) \psi_v \ .
\ee 
Further, using that 
\ba
\bar \psi_{+v} (x) \gamma^\mu \psi_{+v} (x) & = & v^\mu \bar \psi_{+v} (x) \gamma^0 \psi_{+v} (x) \ , \\
\bar \psi_{-v} (x) \gamma^\mu \psi_{-v} (x) & = & {\tilde v}^\mu \bar \psi_{-v} (x) \gamma^0 \psi_{-v} (x) \ , \\
\bar \psi_{\pm v} (x) \gamma^\mu \psi_{ \mp v} (x) & = &  \bar \psi_{\pm v} (x) \gamma^\mu_\perp \psi_{ \mp v} (x) \ ,
\ea
where  $\gamma^\mu_\perp = ( 0 , \gamma^i - v^i {\bf v \cdot \gamma})$, one can write
\begin{widetext}
\ba
\label{generalLag}
{\cal L}_f & = & \sum_{\bf v} \left ( \bar \chi_{+v} (x)  i \gamma^0  v \cdot D \chi_{+v}(x) +  \bar H^1_{-v} (x)    \gamma^0 (2 p + i {\tilde  v} \cdot D ) H^1_{-v}(x) \right. \\
\nonumber
& + & \left. \bar \chi_{+v} (x)  i \slashed{D}_\perp   H^1_{-v}(x) +  \bar H^1_{-v} (x)  i \slashed{D}_\perp   \chi_{+v}(x) \right) 
\nonumber
\\
& + & \sum_{\bf v} \left ( \bar \xi_{-v} (x)  i \gamma^0 {\tilde v} \cdot D \xi_{-v}(x) +  \bar H^2_{+v} (x)    \gamma^0 (-2 p + i   v \cdot D ) H^2_{+v}(x) \right. 
\nonumber
\\
\nonumber
& + & \left. \bar \xi_{-v} (x)  i  \slashed{D}_\perp  H^2_{+v}(x) +  \bar H^2_{+v} (x)  i \slashed{D}_\perp  \xi_{-v}(x) \right)  \ ,
\ea
\end{widetext}
where we have defined $\slashed{D}_\perp = \gamma^\mu_\perp D_\mu$, and we have kept only the terms of the Lagrangian which respect energy-momentum conservation.

It is now possible to integrate out the $H^1_{-v}$ and $H^2_{+v}$ fields, by using their classical equations of motion.
Using 
\ba
(2 p + i {\tilde  v} \cdot D ) H^1_{-v}  + i  \gamma^0 \slashed{D}_\perp \chi_{+v} &=& 0 \ , \\
(- 2 p + i   v \cdot D ) H^2_{+v}  + i  \gamma^0 \slashed{D}_\perp \xi_{-v}& =& 0 \ ,
\ea
one can solve for $H^1_{-v}$
\ba
 H^1_{-v} (x) & = & - \frac{ i \gamma^0}{2 p + i {\tilde  v} \cdot D}  \slashed{D}_\perp \chi_{+v}
 \nonumber
  \\
&  = & -\frac{i \gamma^0}{2 p} \sum_{n=0}^{\infty} \left(- i \frac{  \tilde{v} \cdot D }{2p} \right)^n \slashed{D}_\perp \chi_{+v}(x)  \ ,
\ea
and for $H^2_{+v}$ 
\ba
 H^2_{+v} (x) &=& - \frac{ i \gamma^0}{-2 p + i   v \cdot D}  \slashed{D}_\perp \xi_{-v} \nonumber \\
 &=& \frac{i \gamma^0}{2 p} \sum_{n=0}^{\infty} \left( i \frac{  v \cdot D }{2p} \right)^n \slashed{D}_\perp \xi_{-v}(x)  \ .
\ea

Substituting the lowest-order solutions of $H^1_{-v}, H^2_{+v}$ into the Lagrangian Eq.~(\ref{generalLag}) we end up with
%\begin{widetext}
\ba
{\cal L}_f  & = &  \sum_{\bf v} \left (  \chi^\dagger_{+v} (x)  i   v \cdot D \chi_{+v}(x) -  \chi^\dagger_{+v} (x)  \frac{(\slashed{D}_\perp)^2   }{2p} \chi_{+v}(x) \right)
\nonumber
 \\
 & + &  \sum_{\bf v} \left (  \xi^\dagger_{-v} (x)  i {\tilde  v} \cdot D \xi_{-v}(x) +  \xi^\dagger_{-v} (x)  \frac{(\slashed{D}_\perp)^2   }{2p} \xi_{-v}(x) \right) \ ,
\nonumber
\ea
%\end{widetext}
where fermion and antifermion degrees of freedom are totally decoupled.

The dispersion laws can be derived by taking into account that
\be 
(\slashed{D}_\perp)^2 = (D_\perp)^2 + e {\bf \Sigma} \cdot {\bf v} \, {\bf B} \cdot {\bf v} \ ,
\ee
where $D_\perp = (0, {\bf D} - {\bf v} ({\bf v} \cdot {\bf D}))$. 
Noting that $ \epsilon_q \equiv q^0= p + k_{\parallel} + {\bf k}^2_\perp /(2p) +\cdots$ and that ${\bf q}/ q^0 = {\bf v} + {\bf k}_\perp/p + \cdots$,
 we can then write the dispersion relation for the fermions as
\be
\epsilon_q= q - e \lambda \frac{ {\bf B} \cdot {\bf \hat q}}{2 q} + {\cal O} \left( \frac{1}{q^2} \right) \ ,
\ee
where $\lambda$ is the value of the helicity.
Please note that a similar dispersion law was found for fermions close to the Fermi surface in Ref.~\cite{Son:2012zy} using the HDET,
where in that case $p = \mu$.

In Ref.~\cite{Son:2012zy} the HDET has been used to derive the chiral kinetic theory valid close to the Fermi surface. The fact that the effective field theory we have just derived keeps the same structure as that of HDET, 
both for fermions and for antifermions, suggests that the same methods and results of Ref.~\cite{Son:2012zy} could be used to derive Eq.~(\ref{chiraltreq}) from
a quantum field theoretical framework.  To this aim, one defines a Wigner transform of the two-point function, which depends only on the residual momentum.
After performing a derivative expansion the equation obeyed by the Wigner transform (identified with the classical distribution function)
agrees with Eq.~(\ref{chiraltreq}) when computed to order $\mathcal{O} (1/\mu^2)$. While in  Ref.~\cite{Son:2012zy} the proof required the existence of a Fermi surface, here we note
that it could be extended to more general situations if one uses the OSEFT.

It would be very interesting to check that the OSEFT is able to reproduce the results of different physical quantities which are dominated by on-shell charged fermions in
an electromagnetic relativistic plasma. We leave a further deeper study of this EFT for future work~\cite{future}.

\section{Summary}

We have presented a derivation of the Son and Yamamoto chiral transport approach. The proof is based on performing a diagonalization of the
quantum Dirac Hamiltonian to order $\mathcal{O} (\hbar)$ and deducing from it the leading quantum corrections to the classical equations of motion
and then, to the associated transport theory. In this process, the Berry connection appears modifying the particle's action and also the canonical PB structure
of the theory. While several authors 
realized that a diagonalization procedure can explain the origin of the Berry connection in the particle's action, they forgot that the same
diagonalization procedure leads to a modification of the particle's dispersion law. This is a situation that also occurs in many condensed
matter systems, see e.g. Refs.~\cite{Chang:1995zz,Sundaram:1999zz,Xiao:2005qw}, where a completely analogous correction, proportional to the magnetic field,  emerges.
We would like to recall once more that the modified dispersion relation~(\ref{eq:disp1}, \ref{eq:disp2}) is required, not only for the consistency of the semiclassical approach to ${\cal O} (\hbar)$, but also
to correctly reproduce the anomalous hard thermal loop action  at finite temperature and density~\cite{Son:2012zy,Manuel:2013zaa}.

We have also developed  an effective theory approach for the relevant fermionic degrees of freedom of a plasma that are almost on shell. The OSEFT provides a powerful scheme to study the dynamics of fermions and antifermions on equal footing.
Using the OSEFT  we have computed the fermion dispersion relation in a much simpler way than the semiclassical diagonalization of the Dirac's Hamiltonian, allowing for a systematic expansion to obtain next-to-leading-order corrections. 
Thermal and finite density effects could be studied with it using the standard machinery of thermal quantum field theory. Here, we have only pointed out that it can also be used to derive the chiral transport equation, following
the same line of reasoning  presented in Ref.~\cite{Son:2012zy}. 

 We have also settled the basis for a more general derivation of quantum corrections for massive fermions,
 where the Berry curvature also depends on the mass $m$. The study of the massive case has already started in
Ref.~\cite{Chen:2013iga}, although  in that reference the quantum correction to the dispersion relation~(\ref{eq:ham}) was obviated. In the massive case, an analogous semiclassical transport equation could be derived, after
defining a classical spin vector, together with its classical equations of motion \cite{Bliokh:2005nk}.

{\bf NOTE ADDED:}
After the completion of this manuscript the reference~\cite{Chen:2014cla} appeared, where a modified dispersion relation is obtained for chiral fermions from the path integral, thus correcting the previous diagonalization in~\cite{Stephanov:2012ki}. The dispersion relation is identical as the one derived here for the particular case of $m=0$.

{\bf Acknowledgments:}
We are indebted to  J. Soto  for suggesting to us the possibility of separating particle and antiparticle degrees of freedom using EFT techniques and for a critical reading of the manuscript.
This research was in part supported from Ministerio de Ciencia e Innovaci\'on under contract FPA2010-16963, Programme FP7-PEOPLE-2011-CIG under contract
PCIG09-GA-2011-291679 and Programme TOGETHER from R\'egion Pays de la Loire and the European I3-Hadron Physics programme.

\appendix

\section{Chiral anomaly\label{app:integral}}

We detail the computation of the chiral anomaly in a thermal relativistic plasma. The divergence of the fermion current reads	

\be \pa_\mu j^\mu = e^2 \hbar \lim_{\Delta \rightarrow 0} \int_{S^2(\Delta)} \frac{d{\bf S}}{(2\pi \hbar)^3} \cdot {\bf \Omega} \ f_p \ {\bf E} \cdot {\bf B} \ , \ee
with $f_p$ being the distribution function of the particular fermion species~(\ref{eq:distri1},\ref{eq:distri2}), which already takes into account the modified dispersion laws  Eqs.~(\ref{eq:disp1},\ref{eq:disp2}).

The surface integral is performed around the singular point ${\bf p}=0$ on a two-sphere of radius $\Delta$. Afterwards, the limit $\Delta \rightarrow 0$ should be taken to reach the singular point, where the 
level crossing responsible for the anomaly takes place \cite{Nielsen:1983rb}.
The integral is expressed in spherical coordinates as
\begin{widetext}
\be \pa_\mu j^\mu = e^2 \hbar {\bf E} \cdot {\bf B} \  \lim_{\Delta \rightarrow 0}  \int  \frac{\Delta^2 \sin \theta d\theta d\varphi}{(2\pi \hbar)^3}  \ \frac{\lambda}{2\Delta^2} \left\{ \exp \left[ \frac{1}{T} \left(\Delta - \lambda e \hbar \frac{B  \cos \theta}{2\Delta} \mp \mu \right) \right] +1 \right\}^{-1}  \ , \ee
\end{widetext}
where the sign of the helicity and the chemical potential should be chosen in accordance with the helicity and charge of the fermion.
The azimuthal integral can be trivially done and the polar angle is traded for $x\equiv \cos \theta$:
\begin{widetext}
\be \pa_\mu j^\mu = \pm \frac{e^2}{\hbar^2} \frac{1}{8 \pi^2}  {\bf E} \cdot {\bf B}   \lim_{\Delta \rightarrow 0}  \int_{-1}^1 dx  \left[  \exp \left(  \beta \Delta \mp \gamma \frac{x}{\Delta} \mp \alpha \right) +1  \right]^{-1} \ , \ee
\end{widetext}
where we have called $\beta=1/T$, $\alpha=\mu/T$, and $\gamma=e \hbar B/2T$ for simplicity.

The integral can be exactly done:
\begin{widetext}
\be I_\Delta (\alpha,\beta,\gamma) \equiv \int_{-1}^1 dx  \left[ \exp \left(  \beta \Delta - \gamma \frac{x}{\Delta} - \alpha \right) +1 \right]^{-1} = 2 + \frac{\Delta}{\gamma} \log \left( \frac{1 + e^{ - \alpha + \beta \Delta - \gamma/\Delta}}{1+e^{ - \alpha+\beta \Delta + \gamma/\Delta}} \right) \ , \ee
with the reflection property $I_\Delta (\alpha,\beta,-\gamma)=I_\Delta(\alpha,\beta,\gamma)$.
\end{widetext}
The limit $\Delta\rightarrow 0$ can now be performed:
\be \lim_{\Delta \rightarrow 0} I_\Delta(\pm \alpha, \beta, \pm \gamma) = 1 \ . \ee
We are now ready to combine all fermion species to form the divergence of the axial current:
\begin{widetext}
\be \pa_\mu j_A^\mu = \pa_\mu j_R^\mu - \pa_\mu j_L^\mu= \frac{e^2}{8\pi^2\hbar^2} {\bf E} \cdot {\bf B} \left[ I_{\Delta=0}^R - (-\bar{I}_{\Delta=0}^L) - (-I_{\Delta=0}^L) + \bar{I}_{\Delta=0}^R \right] =\frac{e^2}{2\pi^2 \hbar^2} {\bf E} \cdot {\bf B} \ , \ee
\end{widetext}
where the notation represents the contribution of right-handed fermions, left-handed antifermions, left-handed fermions and right-handed antifermions, in that order.


\begin{thebibliography}{9}
 
 
  
  %\cite{}
\bibitem{Adler} 
  S.~L.~Adler,
  %``Axial vector vertex in spinor electrodynamics,''
  Phys.\ Rev.\  {\bf 177}, 2426 (1969).
  %%CITATION = PHRVA,177,2426;%%

\bibitem{BellJackiw} 
  J.~S.~Bell and R.~Jackiw,
  %``A PCAC puzzle: pi0 --> gamma gamma in the sigma model,''
  Nuovo Cimento \ A {\bf 60}, 47 (1969).
  %%CITATION = NUCIA,A60,47;%%
  
  
  %\cite{Son:2009tf}
\bibitem{Son:2009tf} 
  D.~T.~Son and P.~Surowka,
  %``Hydrodynamics with Triangle Anomalies,''
  Phys.\ Rev.\ Lett.\  {\bf 103}, 191601 (2009)
  [arXiv:0906.5044 [hep-th]].
  %%CITATION = ARXIV:0906.5044;%%
  %195 citations counted in INSPIRE as of 02 Apr 2014
  
  
  %\cite{Kharzeev:2013ffa}
\bibitem{Kharzeev:2013ffa} 
  D.~E.~Kharzeev,
  %``The Chiral Magnetic Effect and Anomaly-Induced Transport,''
  Prog.\ Part.\ Nucl.\ Phys.\  {\bf 75}, 133 (2014)
  [arXiv:1312.3348 [hep-ph]].
  %%CITATION = ARXIV:1312.3348;%%
  %7 citations counted in INSPIRE as of 02 Apr 2014
 
  %\cite{Landsteiner:2012kd}
\bibitem{Landsteiner:2012kd} 
  K.~Landsteiner, E.~Megias and F.~Pena-Benitez,
  %``Anomalous Transport from Kubo Formulae,''
  Lect.\ Notes Phys.\  {\bf 871}, 433 (2013)
  [arXiv:1207.5808 [hep-th]].
  %%CITATION = ARXIV:1207.5808;Stephanov:2012ki%%
  %22 citations counted in INSPIRE as of 04 Apr 2014
  
  %\cite{Son:2012wh}
\bibitem{Son:2012wh} 
  D.~T.~Son and N.~Yamamoto,
  %``Berry Curvature, Triangle Anomalies, and the Chiral Magnetic Effect in Fermi Liquids,''
  Phys.\ Rev.\ Lett.\  {\bf 109}, 181602 (2012)
  [arXiv:1203.2697 [cond-mat.mes-hall]].
  %%CITATION = ARXIV:1203.2697;%%
  
    %\cite{Stephanov:2012ki}
\bibitem{Stephanov:2012ki} 
  M.~A.~Stephanov and Y.~Yin,
  %``Chiral Kinetic Theory,''
  Phys.\ Rev.\ Lett.\  {\bf 109}, 162001 (2012)
  [arXiv:1207.0747 [hep-th]].
  %%CITATION = ARXIV:1207.0747;%%
  

  
%\cite{Son:2012zy}
\bibitem{Son:2012zy} 
  D.~T.~Son and N.~Yamamoto,
  %``Kinetic theory with Berry curvature from quantum field theories,''
  Phys.\ Rev.\ D {\bf 87}, 085016 (2013)
  [arXiv:1210.8158 [hep-th]].
  %%CITATION = ARXIV:1210.8158;%%
  %13 citations counted in INSPIRE as of 14 Oct 2013
  
  

    
    

\bibitem{Chen:2012ca} 
  J.~-W.~Chen, S.~Pu, Q.~Wang and X.~-N.~Wang,
  %``Berry curvature and 4-dimensional monopole in relativistic chiral kinetic equation,''
  Phys.\ Rev.\ Lett.\  {\bf 110}, 262301 (2013)
  [arXiv:1210.8312 [hep-th]].
  %%CITATION = ARXIV:1210.8312;%%
  
%\cite{Manuel:2013zaa}
\bibitem{Manuel:2013zaa} 
  C.~Manuel and J.~M.~Torres-Rincon,
  %``Kinetic theory of chiral relativistic plasmas and energy density of their gauge collective excitations,''
  Phys.\ Rev.\ D {\bf 89}, 096002 (2014)
  [arXiv:1312.1158 [hep-ph]].
  %%CITATION = ARXIV:1312.1158;%%


 
 %\cite{Laine:2005bt}
\bibitem{Laine:2005bt} 
  M.~Laine,
  %``Real-time Chern-Simons term for hypermagnetic fields,''
  J. High Energy Phys. {\bf 0510}, 056 (2005)
  [hep-ph/0508195].
  %%CITATION = HEP-PH/0508195;%%
  %12 citations counted in INSPIRE as of 14 Oct 2013
 
 
%\cite{Foldy:1949wa}
\bibitem{Foldy:1949wa} 
  L.~L.~Foldy and S.~A.~Wouthuysen,
  %``On the Dirac theory of spin 1/2 particle and its nonrelativistic limit,''
  Phys.\ Rev.\  {\bf 78}, 29 (1950).
  %%CITATION = PHRVA,78,29;%%
  %518 citations counted in INSPIRE as of 12 Mar 2014yza

%\cite{Gosselin:2006ht}
\bibitem{Gosselin:2006ht} 
  P.~Gosselin, A.~Berard and H.~Mohrbach,
  %``Semiclassical diagonalization of quantum Hamiltonian and equations of motion with Berry phase corrections,''
  Eur.\ Phys.\ J.\ B {\bf 58}, 137 (2007)
  [hep-th/0603192].
  %%CITATION = HEP-TH/0603192;%%
  %19 citations counted in INSPIRE as of 25 Feb 2014

%\cite{Bliokh:2005nk}
\bibitem{Bliokh:2005nk} 
  K.~Y.~Bliokh,
  %``Topological spin transport of relativistic electron,''
  Europhys.\ Lett.\  {\bf 72}, 7 (2005)
  [quant-ph/0501183].
  %%CITATION = QUANT-PH/0501183;%%
  %21 citations counted in INSPIRE as of 25 Feb 2014



 \bibitem{Berry:1984jv} 
  M.~V.~Berry,
  %``Quantal phase factors accompanying adiabatic changes,''
  Proc.\ R.\ Soc.\ Lond.\ A {\bf 392}, 45 (1984).
  %%CITATION = PRSLA,A392,45;%%  

  %\cite{Jackiw:1987vi}
\bibitem{Jackiw:1987vi} 
  R.~Jackiw,
  %``Three Elaborations on Berry's Connection, Curvature and Phase,''
  Int.\ J.\ Mod.\ Phys.\ A {\bf 03}, 285 (1988).
  %%CITATION = IMPAE,A3,285;%%


%\cite{Chang:2008zza}
\bibitem{Chang:2008zza} 
  M.~C.~Chang and Q.~Niu,
  %``Berry curvature, orbital moment, and effective quantum theory of electrons in electromagnetic fields,''
  J.\ Phys.\ Condens.\ Matter {\bf 20}, 193202 (2008).
  %%CITATION = JCOME,20,193202;%%



  
%\cite{Chen:2013iga}
\bibitem{Chen:2013iga} 
  J.~W.~Chen, J.~y.~Pang, S.~Pu and Q.~Wang,
  %``Kinetic equations for massive Dirac fermions in electromagnetic field with non-Abelian Berry phase,''
  Phys.\ Rev.\ D {\bf 89}, 094003 (2014)
  [arXiv:1312.2032 [hep-th]].
  %%CITATION = ARXIV:1312.2032;%%
  
  
  %\cite{Akamatsu:2014yza}
\bibitem{Akamatsu:2014yza} 
  Y.~Akamatsu and N.~Yamamoto,
  %``Chiral Langevin theory for non-Abelian plasmas,''
  arXiv:1402.4174 [hep-th].
  %%CITATION = ARXIV:1402.4174;%%
  %1 citations counted in INSPIRE as of 07 Mar 2014yza




%\cite{Duval:2005vn}
\bibitem{Duval:2005vn} 
  C.~Duval, Z.~Horvath, P.~A.~Horvathy, L.~Martina and P.~Stichel,
  %``Berry phase correction to electron density in solids and 'exotic' dynamics,''
  Mod.\ Phys.\ Lett.\ B {\bf 20}, 373 (2006)
  [cond-mat/0506051].
  %%CITATION = COND-MAT/0506051;%%
  %35 citations counted in INSPIRE as of 26 Mar 2014

  
  
 
 
%\cite{Morrison:1998zz}
\bibitem{Morrison:1998zz} 
  P.~J.~Morrison,
  %``Hamiltonian description of the ideal fluid,''
  Rev.\ Mod.\ Phys.\  {\bf 70}, 467 (1998).
  %%CITATION = RMPHA,70,467;%%
  %21 citations counted in INSPIRE as of 26 Mar 2014yza
  

\bibitem{future}
In preparation.  
  
  
 %\cite{Charles:1998dr}
\bibitem{Charles:1998dr} 
  J.~Charles, A.~Le Yaouanc, L.~Oliver, O.~Pene and J.~C.~Raynal,
  %``Heavy to light form-factors in the heavy mass to large energy limit of QCD,''
  Phys.\ Rev.\ D {\bf 60}, 014001 (1999)
  [hep-ph/9812358].
  %%CITATION = HEP-PH/9812358;%%
  %292 citations counted in INSPIRE as of 05 Feb 2014

%\cite{Hong:1998tn}
\bibitem{Hong:1998tn} 
  D.~K.~Hong,
  %``An Effective field theory of QCD at high density,''
  Phys.\ Lett.\ B {\bf 473}, 118 (2000)
  [hep-ph/9812510].
  %%CITATION = HEP-PH/9812510;%%
  
  %\cite{Hong:1999ru}
\bibitem{Hong:1999ru} 
  D.~K.~Hong,
  %``Aspects of high density effective theory in QCD,''
  Nucl.\ Phys.\ B {\bf 582}, 451 (2000)
  [hep-ph/9905523].  
  




%\cite{Chang:1995zz}
\bibitem{Chang:1995zz} 
  M.~-C.~Chang and Q.~Niu,
  %``Berry Phase, Hyperorbits, and the Hofstadter Spectrum,''
  Phys.\ Rev.\ Lett.\  {\bf 75}, 1348 (1995).
  %%CITATION = PRLTA,75,1348;%%
  %32 citations counted in INSPIRE as of 16 Dec 2013

%\cite{Sundaram:1999zz}
\bibitem{Sundaram:1999zz} 
  G.~Sundaram and Q.~Niu,
  %``Wave-packet dynamics in slowly perturbed crystals: Gradient corrections and Berry-phase effects,''
  Phys.\ Rev.\ B {\bf 59}, 14915 (1999).
  %%CITATION = PHRVA,B59,14915;%%	
  %50 citations counted in INSPIRE as of 16 Dec 2013


%\cite{Xiao:2005qw}
\bibitem{Xiao:2005qw} 
  D.~Xiao, J.~-r.~Shi and Q.~Niu,
  %``Berry phase correction to electron density of states in solids,''
  Phys.\ Rev.\ Lett.\  {\bf 95}, 137204 (2005)
  [cond-mat/0502340].
  %%CITATION = COND-MAT/0502340;%%
  %41 citations counted in INSPIRE as of 16 Dec 2013

  %\cite{Chen:2014cla}
\bibitem{Chen:2014cla} 
  J.~-Y.~Chen, D.~T.~Son, M.~A.~Stephanov, H.~-U.~Yee, and Y.~Yin,
  %``Lorentz Invariance in Chiral Kinetic Theory,''
  arXiv:1404.5963 [hep-th].
  %%CITATION = ARXIV:1404.5963;%%


  \bibitem{Nielsen:1983rb} 
  H.~B.~Nielsen and M.~Ninomiya,
  %``Adler-bell-jackiw Anomaly And Weyl Fermions In Crystal,''
  Phys.\ Lett.\ B {\bf 130}, 389 (1983).
  %%CITATION = PHLTA,B130,389;%%

  




  \end{thebibliography}
\end{document}